\documentclass[sort&compress,5p]{elsarticle}

\journal{SoftwareX}

\usepackage{bm}
\newcommand{\ve}[1]{\bm{#1}}

\newcommand{\bF}{\ve{F}}

\newcommand{\dd}{\text{d}}

\newcommand{\Rdot}{\dot{R}}
\newcommand{\Rddot}{\ddot{R}}

\newcommand{\bxi}{\boldsymbol{\xi}}

\newcommand{\eps}{\varepsilon}

\newcommand{\hw}{\widehat{w}}
\newcommand{\hxi}{\widehat{\xi}}

\newcommand{\bhw}{\bm{\widehat{w}}}
\newcommand{\bhxi}{\boldsymbol{\widehat{\xi}}}

\newcommand{\vbk}{\vec{\ve{k}}}

\newcommand{\vbm}{\vec{\ve{M}}}

\newcommand{\vbF}{\vec{\ve{F}}}

\newcommand{\mom}{M}

\newcommand\Rey{\mbox{Re}}  

\usepackage{pdfpages}
\makeatletter
\AtBeginDocument{\let\LS@rot\@undefined}
\makeatother

\usepackage{epsf}
\usepackage{amsmath}
\usepackage{amssymb}
\usepackage{stmaryrd}
\usepackage{setspace}
\usepackage{enumerate}
\usepackage{graphicx}
\usepackage{float}
\usepackage{footnote}
\usepackage{microtype}
\usepackage{scalefnt}
\usepackage{microtype}
\usepackage{transparent}
\usepackage{array}
\usepackage{siunitx}
\usepackage{xcolor}
\usepackage[]{algorithm2e}
\usepackage{listings}
\usepackage{arydshln}
\usepackage{lmodern}
\usepackage{upquote}
\usepackage{mmacells}
\usepackage{enumitem}

\mmaDefineMathReplacement[≤]{<=}{\leq}
\mmaDefineMathReplacement[≥]{>=}{\geq}
\mmaDefineMathReplacement[≠]{!=}{\neq}
\mmaDefineMathReplacement[→]{->}{\to}[2]
\mmaDefineMathReplacement[⧴]{:>}{:\hspace{-.2em}\to}[2]
\mmaDefineMathReplacement{∉}{\notin}
\mmaDefineMathReplacement{∞}{\infty}
\mmaDefineMathReplacement{𝕕}{\mathbbm{d}}

\mmaSet{
  morelst={alsoletter={`}},
  morefv={gobble=2},
  moredefined={Image,ImageData,img,na},
}

\newcommand\LL[1]{\multicolumn{1}{|c}{#1}}
\newcommand\RR[1]{\multicolumn{1}{c|}{#1}}
\newcommand\LR[1]{\multicolumn{1}{|c|}{#1}}

\newcommand\LLd[1]{\multicolumn{1}{:c}{#1}}
\newcommand\RRd[1]{\multicolumn{1}{c:}{#1}}

\begin{document}

\begin{frontmatter}

\title{QBMMlib: A library of quadrature-based moment methods}

\author[add1]{Spencer H. Bryngelson}
\ead{spencer@caltech.edu}
\author[add1]{Tim Colonius}
\author[add2,add3]{Rodney O. Fox}

\address[add1]{Division of Engineering and Applied Science, 
    California Institute of Technology,
    Pasadena, CA 91125, USA}
\address[add2]{Department of Chemical and Biological Engineering, Iowa State University, Ames, IA 50011, USA}
\address[add3]{Center for Multiphase Flow Research and Education, Iowa State University, Ames, IA 50011, USA}

\begin{abstract}
    QBMMlib is an open source Mathematica package of quadrature-based moment methods and their algorithms.
    Such methods are commonly used to solve fully-coupled disperse flow and combustion problems, though formulating and closing the corresponding governing equations can be complex.
    QBMMlib aims to make analyzing these techniques simple and more accessible.
    Its routines use symbolic manipulation to formulate the moment transport equations for a population balance equation and a prescribed dynamical system.
    However, the resulting moment transport equations are unclosed.
    QBMMlib trades the moments for a set of quadrature points and weights via an inversion algorithm, of which several are available.
    Quadratures then closes the moment transport equations.
    Embedded code snippets show how to use QBMMlib, with the algorithm initialization and solution spanning just 13 total lines of code.
    Examples are shown and analyzed for linear harmonic oscillator and bubble dynamics problems.
\end{abstract}


\end{frontmatter}

\begin{table}[h!]
    \centering
        \footnotesize
    \begin{tabular}{ l l }
        \hline \hline
        Version & v1.0 \\ 
        Link to code & \href{https://github.com/sbryngelson/QBMMlib}{github.com/sbryngelson/QBMMlib} \\ 
        License   & GPL 3\\
        Versioning & git \\ 
        Language & Wolfram Language / Mathematica \\
        Requirements & Mathematica v8.0+ \\ 
        Support email & \href{mailto:spencer@caltech.edu}{spencer@caltech.edu} \\ \hline \hline
    \end{tabular}
    \caption{Code metadata}
\end{table}

\section{Motivation and significance}\label{s:intro}

QBMMlib is an open-source library and solves population balance equations (PBEs) using quadrature-based moment methods (QBMMs).
PBEs model the evolution of a number density function (NDF)~\citep{ramkrishna00,chapman1990,vanni00,smoluchowski1916,solsvik2015}.
Such models are useful, for example, in fluid dynamics simulations involving dispersions, wherein the NDF evolution can represent growth, shrinkage, coalescence, breakup, and relative motion~\citep{buffo2012,Buffo2013,liao14,li2017,gao2016,fox2008,desjardins08,nguyen2016,kong2017b},
Example engineering applications of this are combustion (e.g.\ soot dynamics in flames)~\citep{Kazakov1998,Balthasar2003,Pedel2014,Mueller2009} and aerosols (e.g.\ sprays)~\citep{sibra2017,laurent2001,hussain2015}.

\begin{figure*}
    \centering
    \includegraphics{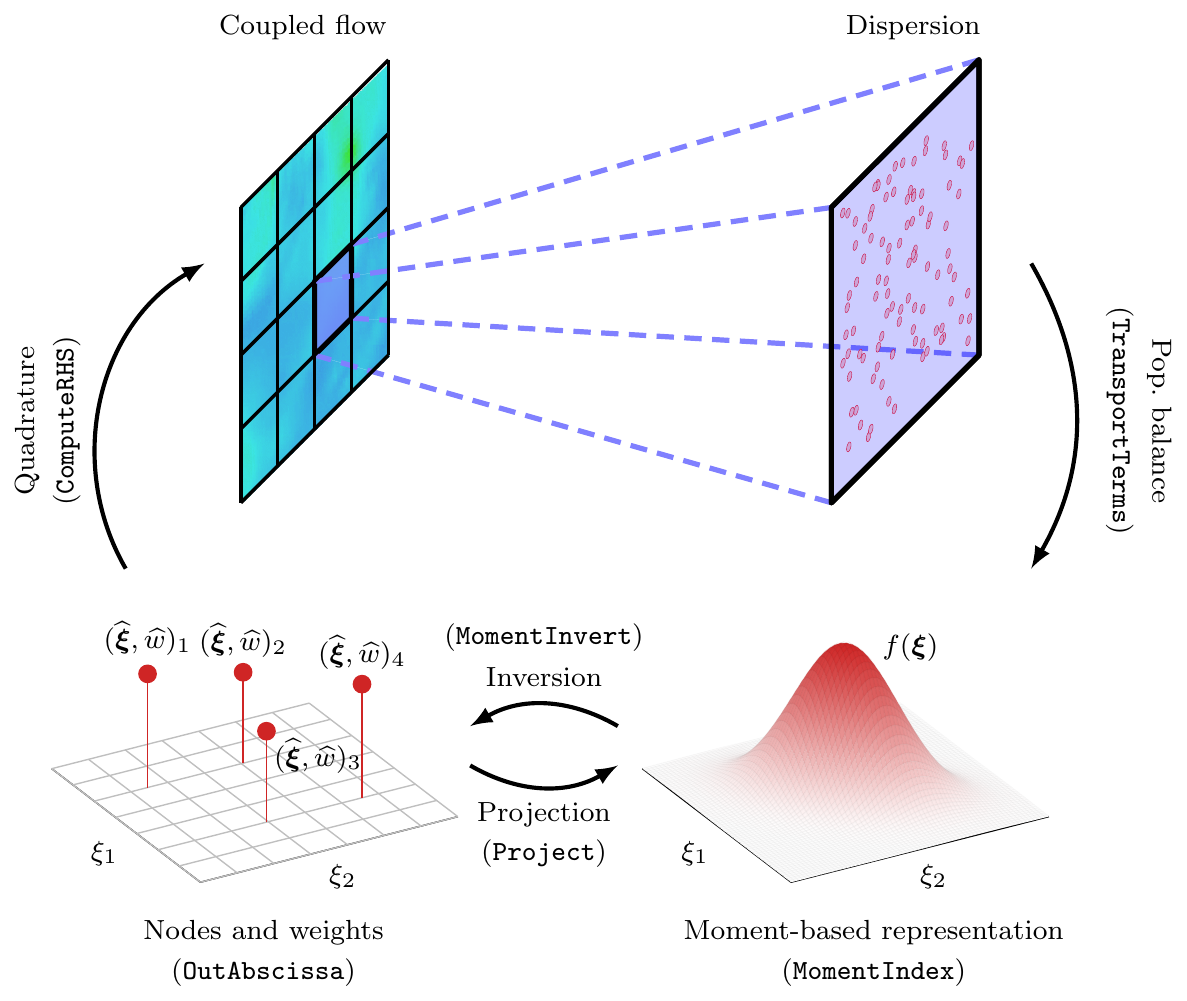}
    \caption{Schematic illustration of the QMOM solution method for flowing dispersions.
        QBMMlib routines are in parentheses.}
    \label{f:schematic}
\end{figure*}

PBEs can be solved by the method of classes~\citep{ando11,bryngelson19} or the method of moments (MOM)~\citep{hulburt64,moyal49}.
QBMMlib employs the MOM because it can more naturally handle problems with multiple internal coordinates (e.g.\ velocities).
Figure~\ref{f:schematic} shows a typical QBMM-based solution procedure.
The MOM represents the NDF via a set of statistical moments and the transport equations for them follow from the PBE.
Inverting the moments to a set of weights and abscissas provides a basis for approximating the unclosed transport equations via quadrature (QMOM)~\citep{mcgraw97}.

Variations on QMOM are plentiful.
One can change the inversion procedure: Wheeler's algorithm can solve single internal coordinate problems~\citep{Wheeler1974} and algorithms exist for enforcing distribution shape (extended-QMOM (EQMOM)~\citep{yuan12}, anisotropic-Gaussian~\citep{patel17,kong17}) and hyperbolicity (hyperbolic-QMOM (HyQMOM)~\citep{Fox2018}).
The quadrature weights and abscissas can also evolve directly (direct-QMOM (DQMOM)~\citep{marchisio05,fox03,fox06}).
One complication is that multiple internal coordinate problems do not admit a unique choice in moment set~\citep{marchisio13}.
However, conditioning one direction on the others provides a particularly robust moment inversion technique (conditional-QMOM (CQMOM)~\citep{yuan11} and -HyQMOM (CHyQMOM)~\citep{patel19}).
For these reasons, QBMMlib uses Wheeler's algorithm (or its adaptive counterpart) or HyQMOM for one-dimensional moment inversion and CQMOM and CHyQMOM handle multi-dimensional problems.

There is one other actively developed open source QBMM solver: OpenQBMM~\citep{Passalacqua2018,passalacqua19}.
It is a library for OpenFOAM~\cite{weller98} and implements CQMOM and (3-node) CHyQMOM.
MFiX~\citep{fan2004,marchisio05} and Fluidity~\citep{davies2011} use DQMOM, though modern conditional methods (e.g.\ CQMOM and CHyQMOM) generally outperform it~\citep{Buffo2013}.
Note that these are fully-coupled flow solvers.
QBMMlib instead decouples these problems and solves the moment transport equations directly for an input dynamical system.
This makes it preferable for prototyping and testing on novel physical problems.
In pursuit of this, QBMMlib places emphasis on expressive programming, simple interfaces, and symbolic computation where possible.
As a result, it can solve PBE-based problems with just a few lines of code.

Section~\ref{s:overview} describes QBMMlib's implementation of the PBE and QBMMs.
Section~\ref{s:examples} verifies its methods and demonstrates its capabilities for three example problems.
Section~\ref{s:impact} discusses the utility and novelty of QBMMlib, which concludes the paper.

\section{Software description}\label{s:overview}

QBMMlib is a collection of Mathematica functions for solving PBEs via QBMMs.
Table~\ref{t:routines} describes the public-facing routines.
These routines are also documented and accessible in Mathematica via
\begin{mmaCell}[moredefined={QBMMlib`}]{Code}
  Get["QBMMlib"];
  ?QBMMlib`*
\end{mmaCell}
Figure~\ref{f:schematic} illustrates their places in the model and its solution procedure.
\texttt{TransportTerms} computes the moment transport equations for the moment set of \texttt{MomentIndex}.
\texttt{MomentInvert} inverts these moments to weights and abscissas that close the moment set via quadrature (\texttt{ComputeRHS}) and project it onto a realizable moment space (\texttt{Project}).
The next sections describe the details of these routines.

\begin{table*}
	\centering
    \footnotesize
    \begin{tabular}{   l@{\hskip 0.2in} r@{\hskip 0.1in} l }
        \hline \hline 
        \texttt{TransportTerms} & 	Input:       & Governing equation (\texttt{eqn}, e.g.~\eqref{e:gov}) and its variables \\
                                &   Output:      & Coefficients (\texttt{coefs}) and exponents (\texttt{exps}) of moment transport equations  \vspace{0.2cm}\\

        \texttt{MomentIndex}    & 	Input:       & Number of nodes (\texttt{n}, $N_{\bhxi}$), inversion method (\texttt{method})   \\
                                &   Output:      & Moment set indices (\texttt{momidx}, $\vbk$)  \\
                                &   Options:     & Number of permutations (default: \texttt{1}) \vspace{0.2cm}  \\ 

        \texttt{MomentInvert}   & 	Input:       & Moment set (\texttt{moments}, $\vbm$) and its indices  \\ 
                                &   Output:      & Optimal set of abscissas (\texttt{xi}, $\bhxi$) and weights (\texttt{w}, $\bhw$)  \\
                                &   Options:     & Method, Permutation (default: \texttt{12} ($\xi_1 | \xi_2$), $N_{\bxi} >1$ only) \vspace{0.2cm}  \\ 

        \texttt{ComputeRHS}     & 	Input:       & Abscissas, weights, moment set indices, transport coefficients  \\
                                &   Output:      & Right-hand-side of moment transport equation (\texttt{rhs}, $\vbF$) \\
                                &   Options:     & Third coordinate direction abscissas  ($N_{\bxi} = 2$ only) \vspace{0.2cm} \\

        \texttt{Project}       	& 	Input:       & Abscissas, weights, moment set indices \\  
                                &   Output:      & Projected moment set (\texttt{momentsP}, $\vbm$) \vspace{0.2cm} \\


        \texttt{OutAbscissa}	& 	Input:       & Abscissas \\  
                                &   Output:      & Threaded abscissas\\

		\hline \hline
	\end{tabular}
    \caption{Example public-facing routines. Parenthetical variables correspond to the code snippets and notation of section~\ref{s:overview}.}
	\label{t:routines}
\end{table*}

\subsection{Population balance equations $\mathrm{(}\mathtt{TransportTerms}\mathrm{)}$}\label{s:pbe}

QBMMlib can solve one- and two-dimensional populations balance equations.
A third direction can be added if its NDF is stationary.
The two-dimensional case is detailed here without loss of generality. 
For illustration, consider
\begin{gather}
    \ddot{x} = g(x,\dot{x}),
    \label{e:gov}
\end{gather}
where the dots indicate partial time derivatives, $g$ is a function, and $\bxi = \{ x, \dot{x} \}$ are the internal coordinates.
The number density function $f$ describes the state and statistics of this system in the $\bxi$-space.
A population balance equation (PBE) governs $f$ as
\begin{gather}
    \frac{ \partial f}{\partial t} + 
    \frac{\partial}{\partial x} (f \dot{x} ) + 
    \frac{\partial}{\partial \dot{x}} (f \ddot{x} ) = 0,
    \label{e:master}
\end{gather}
where the zero right-hand-side indicates conservation of $f$, though sinks and sources can model aggregation and breakup~\citep{marchisio13}.
Quadrature-based methods to solving~\eqref{e:master} represent $f$ by a set of raw moments $\vbm$ as $f(\vbm)$.
The moment indices $\vbk = \{l,m\} = \{ (0,0), \dots \}$ associated the carried moment set $\vbm = \{ M_{l,m} \} $ depend upon the moment inversion procedure and the number of quadrature points (details follow in section~\ref{s:inversion}).

The raw moments are
\begin{gather}
    \mom_{p_1,\dots,p_{N_{\bxi}}} \equiv \int_\Omega f(\bxi) \prod_{j=1}^{N_{\bxi}}  \xi_{j}^{p_j} \dd \xi_j
    \label{e:moment}
\end{gather}
where $p_j$ (for $j = 1,\dots,N_{\bxi}$) are the moment indices, $N_{\bxi}$ is the number of internal coordinates ($N_{\bxi} = 2$ in~\eqref{e:gov}), and $\Omega$ is the domain of $f$.
These moments evolve as
\begin{gather}
    \frac{\partial \vbm}{\partial t} = \vbF ( \vbm ),
	\label{e:system}
\end{gather}
where, for~\eqref{e:gov}, 
\begin{gather}
    F_{l,m} = l \mom_{l-1,m+1} + m \int_\Omega \ddot{x} \,  x^l \dot{x}^{m-1} f \, \dd \bxi.
	\label{e:rhs}
\end{gather}
This forcing follows from the PBE via integration-by-parts~\citep{Bryngelson2020}.
For the prescribed dynamics $\ddot{x}$ (as in~\eqref{e:gov}), the integral term of~\eqref{e:rhs} is equivalent to a sum of moments.
For example, if
\begin{align}
    \ddot{x} = x + \dot{x}, & \nonumber\\ 
    \quad\text{then}&\quad
    \int_\Omega \ddot{x} x^l \dot{x}^{m-1} f \dd \bxi = \mom_{l+1,m-1} + \mom_{l,m},  \nonumber\\
    \quad\text{so}&\quad
    F_{l,m} = l \mom_{l-1,m+1} + m (\mom_{l+1,m-1} + \mom_{l,m}).
    \label{e:exrhs}
\end{align}
The routine \texttt{TransportTerms} manipulates the PBE (as in~\ref{e:exrhs}) to determine the coefficients and moment indices that constitute $\bF$.
The code snippet below demonstrates this functionality for the dynamics of~\ref{e:gov}.
\begin{mmaCell}[moredefined={var,eqn,coefs,exps,TransportTerms}]{Code}
  eqn = x[t]+x'[t] == x''[t]; 
  {coefs,exps} = 
       TransportTerms[eqn,x[t],t]
\end{mmaCell}
\begin{mmaCell}[]{Output}
  \{\{c[2],c[2],c[1]\},\{\{1+c[1],-1+c[2]\},
   \{c[1],c[2]\},\{-1+c[1],1+c[2]\}\}\}
\end{mmaCell}
Here, the unassigned coefficients \texttt{c[1]} and \texttt{c[2]} correspond to the moment indices $l$ and $m$ of~\ref{e:rhs}.

\subsection{Moment inversion and quadrature weights \\ $\mathrm{(}\mathtt{MomentIndex,MomentInvert}\mathrm{)}$}\label{s:inversion}

\texttt{MomentInvert} inverts the set of raw moments $\vbm$ into a set of quadrature weights $\bhw$ and abscissas (nodes) $\bhxi$: 
\begin{gather}
    \vbm \to \{ \bhw, \bhxi \}
    \label{e:mominv}
\end{gather}
Many algorithms can perform this procedure, each with its own relative merits, as discussed in section~\ref{s:intro}.
Common approaches for one-dimensional moment sets ($N_{\bxi} = 1$) are QMOM~\citep{hulburt64,mcgraw97} and hyperbolic QMOM (HyQMOM)~\citep{Fox2018}.
For higher-dimensional moment sets ($N_{\bxi} > 1$) conditioned moment methods are cheaper and more stable than performing QMOM on each coordinate direction individually~\citep{yuan11}.
Such conditioned methods perform 1D moment inversion in one coordinate direction, then condition the next directions on the previous ones~\citep{yuan11}.
Examples of these are conditional-QMOM (CQMOM) and conditional-HyQMOM (CHyQMOM)~\citep{Fox2018,patel19}.
The order that this conditioning is done is called the permutation.
For 2D problems their are two permutations, $\xi_1 | \xi_2$ (coordinate direction $\xi_2$ conditioned on $\xi_1$) and the reverse, $\xi_2 | \xi_1$.

\renewcommand{\arraystretch}{1.25}
\setlength{\tabcolsep}{4pt}
\begin{table*}
    \caption{Example optimal $N_{\bxi} = 2$ moment sets for permutations as labeled.}
    \centering
    {\footnotesize
    \begin{tabular}{ c@{\hskip 10pt} l c l }
        & (a) CQMOM [----- $\xi_1|\xi_2$ -- -- $\xi_2|\xi_1$] & & (b) CHyQMOM \vspace{-0.25cm} \\
        \raisebox{-1.25cm}{\rotatebox[origin=c]{90}{(i) $N_{\hxi} = 2$}} & 
        \begin{tabular}[t]{ c c c c }
            \hline \hline
            $\mom_{0,0}$ & $\mom_{0,1}$ & $\mom_{0,2}$ & $\mom_{0,3}$ \\ \cdashline{3-4}
            $\mom_{1,0}$ & $\mom_{1,1}$ & \LLd{$\mom_{1,2}$}  & \RRd{$\mom_{1,3}$} \\ \cdashline{3-4}\cline{2-2}
            $\mom_{2,0}$ & \LR{$\mom_{2,1}$} &  & \\
            $\mom_{3,0}$ & \LR{$\mom_{3,1}$} &  & \\
            \hline \hline
        \end{tabular}
        &
        \phantom{11111}
        &
        \begin{tabular}[t]{ c c c c }
            \hline \hline
            $\mom_{0,0}$ & $\mom_{0,1}$ & $\mom_{0,2}$ & \phantom{$\mom_{0,3}$}  \\
            $\mom_{1,0}$ & $\mom_{1,1}$ & & \\
            $\mom_{2,0}$ & & & \\
             & & & \\
            \hline \hline
        \end{tabular} \\
            \raisebox{-1.85cm}{\rotatebox[origin=c]{90}{(ii) $N_{\hxi} = 3$}} & 
        \begin{tabular}[t]{ c c c c c c }
            \hline \hline 
            $\mom_{0,0}$ & $\mom_{0,1}$ & $\mom_{0,2}$ & $\mom_{0,3}$ & $\mom_{0,4}$ & $\mom_{0,5}$ \\\cdashline{4-6}
            $\mom_{1,0}$ & $\mom_{1,1}$ & $\mom_{1,2}$ & \LLd{$\mom_{1,3}$} & $\mom_{1,4}$ & \RRd{$\mom_{1,5}$} \\ 
            $\mom_{2,0}$ & $\mom_{2,1}$ & $\mom_{2,2}$ & \LLd{$\mom_{2,3}$} & $\mom_{2,4}$ & \RRd{$\mom_{2,5}$} \\ \cdashline{4-6}\cline{2-3}
            $\mom_{3,0}$ & \LL{$\mom_{3,1}$} & \RR{$\mom_{3,2}$} & & & \\
            $\mom_{4,0}$ & \LL{$\mom_{4,1}$} & \RR{$\mom_{4,2}$} & & & \\
            $\mom_{5,0}$ & \LL{$\mom_{5,1}$} & \RR{$\mom_{5,2}$} & & & \\
            \hline \hline 
        \end{tabular}
        & &
        \begin{tabular}[t]{ c c c c c c }
            \hline \hline 
            $\mom_{0,0}$ & $\mom_{0,1}$ & $\mom_{0,2}$ & $\mom_{0,3}$ & $\mom_{0,4}$ & \phantom{$\mom_{0,5}$} \\
            $\mom_{1,0}$ & $\mom_{1,1}$ & & & &  \\
            $\mom_{2,0}$ & & & & & \\
            $\mom_{3,0}$ & & & & & \\
            $\mom_{4,0}$ & & & & & \\
             & & & & & \\
            \hline \hline 
        \end{tabular} 
    \end{tabular}
    }
    \label{t:momset}
\end{table*}

The indices that makeup a so-called optimal set $\vbm$ depend on the moment inversion method and the number of quadrature nodes used in each internal coordinate direction $N_{\hxi}$.
Here, ``optimal'' constrains the number of moments (and their order) required to yield a full-rank and square coefficient matrix~\citep{fox09_2}.
Optimal moment sets are more stable and smaller (and so cheaper) than non-optimal ones.
For $N_{\bxi} = 1$ single-coordinate problems the optimal moment indices are $\vbm = \{ \mom_0,\mom_1,\dots,\mom_{2 N_{\hxi} -1} \}$ for QMOM and $\vbm = \{ \mom_0,\mom_1,\dots,\mom_{2 N_{\hxi} -2} \}$ for HyQMOM.
Table~\ref{t:momset} shows the optimal moment sets QBMMlib uses for $N_{\bxi} = 2$ two-dimensional problems~\citep{yuan11,Fox2018}.
\texttt{MomentIndex} computes these moment indices given the inversion algorithm (\texttt{method}) and number of quadrature points (\texttt{n}) in each internal coordinate direction ($N_{\hxi}$).
The code snippet below computes the moment set corresponding to $N_{\hxi_1} = N_{\hxi_2} = 2$ via CHyQMOM.
\begin{mmaCell}[moredefined={n,method,momidx,MomentIndex},functionlocal=i]{Code}
  method = "CHYQMOM";
  n = {2,2};
  momidx = MomentIndex[n,method]
\end{mmaCell}
\begin{mmaCell}{Output}
  \{\{0,0\},\{1,0\},\{0,1\},\{2,0\},\{1,1\},\{0,2\}\}
\end{mmaCell}

\texttt{MomentInvert} then inverts the moment set $\vbm$ to a set of weights and abscissas (as in~\eqref{e:mominv}).
In the following Mathematica code snippet, the moment set $\vbm$ (\texttt{moments}) is initialized via a two-dimensional Gaussian distribution (\texttt{BinormalDistribution}), though in principle $\vbm$ can be any realizable moment set.
The \texttt{method} inversion algorithm then converts it to quadrature weights (\texttt{w}) and abscissas (\texttt{xi}).
\begin{mmaCell}[moredefined={w,xi,MomentInvert,method,f,BinormalDistribution,mu1,mu2,sig1,sig2,rho,moments,Moment,GenMoment,momidx},pattern={i_,i}]{Code}
  mu1 = mu2 = 1; sig1 = sig2 = 0.3; 
  rho = 0.5;
  f = BinormalDistribution[{mu1,mu2},
        {sig1,sig2},rho];
  GenMoment[i_] := Moment[f,i];
  moments = Map[GenMoment,momidx];
  {w,xi} = MomentInvert[moments,momidx,
       Method->method];
\end{mmaCell}

\begin{figure}
    \centering
    \includegraphics{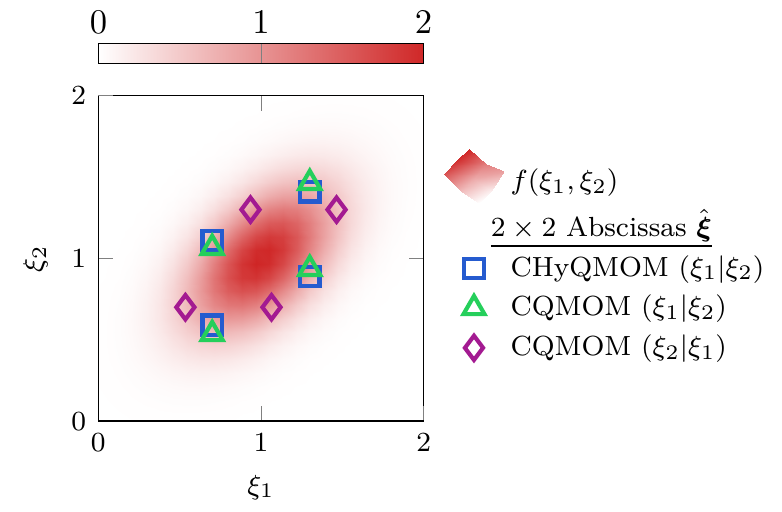}
    \caption{Abscissas $\bhxi$ corresponding to the number density function example $f$ of section~\ref{s:inversion}.
    Different moment inversion algorithms and permutations are shown for $N_{\hxi_1} = N_{\hxi_2} = 2$ as labeled.}
    \label{f:abscissa}
\end{figure}

Figure~\ref{f:abscissa} shows the abscissas for different moment inversion algorithms.
Their locations in the internal coordinate space $\bxi$ are different, though their weights are such that each quadrature reproduces the exact (up to) second-order moments.
We verify that QBMMlib has this property to the carried precision. 
We discuss the QBMMlib quadrature routines used for this next.

\subsection{Moment system closure via quadrature $\mathrm{(}\mathtt{ComputeRHS}\mathrm{)}$}\label{s:quadrature}

Quadrature approximates the raw moments defined in~\eqref{e:moment} and required by~\eqref{e:rhs} as
\begin{gather}
    \prod_{j=1}^{N_{\bxi}} \sum_{i=1}^{N_{\hxi_j}} \hw_{j,i}  \, \hxi_{j,i}^{p_j} \to \mom_{p_1,\dots,p_{N_{\bxi}}}
\end{gather}
where $\hxi_{j,i}$ (for $i = 1,\dots,N_{\hxi_j}$) are the abscissa for internal coordinate direction $\xi_j$ (for $j = 1,\dots,N_{\bxi}$).
These quadrature approximations build $\vbF$ of~\eqref{e:rhs} (\texttt{F}) via the QBMM function \texttt{ComputeRHS}.
\texttt{ComputeRHS} approximates (via quadrature) and sums the required moments (\texttt{exps}) and their coefficients (\texttt{coefs}) for each moment index (\texttt{momidx}). 
The code snippet below shows this implementation in QBMMlib.
\begin{mmaCell}[moredefined={F,MomentInvert,ComputeRHS,rhs,method,moments,w,xi,momidx,exps,coefs},functionlocal={i,j}]{Code}
  F = ComputeRHS[w,xi,momidx,
                 {coefs,exps}];
\end{mmaCell}

\subsection{Realizable time integration $\mathrm{(}\mathtt{Project}\mathrm{)}$}

Stable and realizable time integration of~\eqref{e:system} requires recasting the moment set $\vbm$ from its weights and abscissas~\citep{nguyen2016}.
QBMMlib function \texttt{Project} performs this projection.
A time integrator (e.g.\ Euler's method) then computes the next iteration of the moment set (\texttt{moments}).
The code snippet below shows an example QBMMlib projection and Euler time step (with time step size \texttt{dt}).
\begin{mmaCell}[moredefined={F,w,xi,momidx,dt,Project,momentsP,moments},functionlocal={i,j}]{Code}
  momentsP = Project[w,xi,momidx];
  moments = momentsP + dt F;
\end{mmaCell}
QBMMlib also includes an adaptive strong-stability-preserving (SSP) third-order-accurate Runge--Kutta (RK) time integrator, \texttt{RK23}~\citep{gottlieb11}.
The difference between the SSP--RK3 solution an embedded second-order-accurate SSP-RK solution provides a first-order approximation of the time step error.
The time step size adjustment is then proportional to this error.
The illustrative examples of the next section use this adaptive time stepping procedure. 

\section{Illustrative examples}\label{s:examples}

\subsection{Linear harmonic oscillator}

The example case of section~\ref{s:overview}, including~\ref{e:exrhs}, is a linear harmonic oscillator.
This moment system is linear and thus closed, so it can also verify the solution methods of QBMMlib via comparison to Monte Carlo simulations.

\begin{figure}
    \centering
    \includegraphics{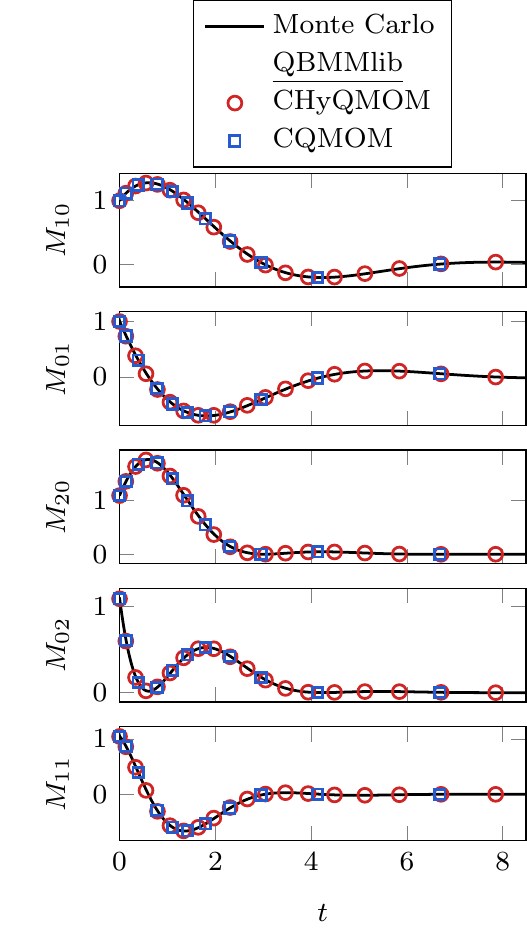}
    \caption{Evolution of the first- and second-order moments in time $t$ for the linear harmonic oscillator example problem. 
    Monte Carlo simulation serves as the truth and the symbols show QBMMlib solutions using $N_{\hxi_1} = N_{\hxi_2} = 2$ CHyQMOM and CQMOM.}
    \label{f:verif}
\end{figure}

\begin{figure}
    \centering
    \includegraphics{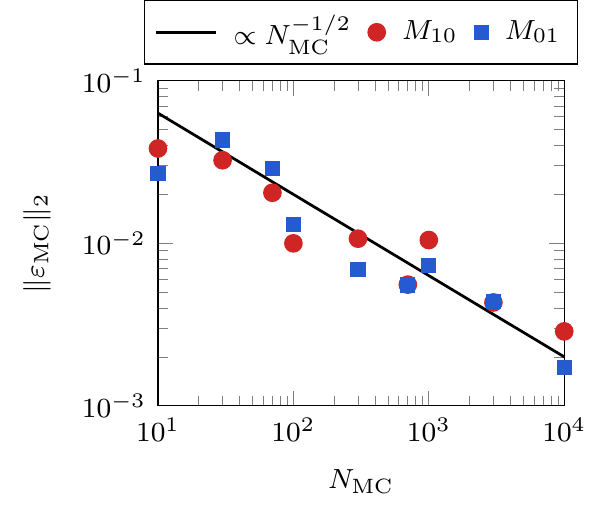}
    \caption{Nominal differences $\lVert \eps_\text{MC} \rVert_2$ between the first-order moments of Monte Carlo simulation ensembles (of size $N_\text{MC}$) and the approximations of CHyQMOM (via QBMMlib).
    The expected convergence power law $\lVert \eps_\text{MC} \rVert_2 \propto 1/\sqrt{N_\text{MC}}$ is also shown.}
    \label{f:converg}
\end{figure}

Figure~\ref{f:verif} shows the evolution of CQMOM and CHyQMOM moment sets and compares them to Monte Carlo surrogate truth solutions.
The behavior of the first-order moments ($\mom_{10}$ and $\mom_{01}$) match the positions and velocities expected of a linear oscillator.
Further, the QBMMlib solutions match the moments of the Monte Carlo simulations to plotting accuracy.
The $L_2$ norm $\lVert \cdot \rVert_2$ of the error quantifies this matching as
\begin{gather}
     \eps_\text{MC}(t) \equiv \frac{\mom_{ij}^\text{(MC)}(t) - \mom_{ij}^\text{(QBMM)}(t)}{\max_t \mom_{ij}^\text{(QBMM)}(t)}
\end{gather}
where superscripts (MC) and (QMOM) are correspond to Monte Carlo and QBMMlib simulations, respectively.
Figure~\ref{f:converg} shows $\lVert \eps_\text{MC} \rVert_2$ for varying Monte Carlo ensemble size $N_\text{MC}$ and QBMMlib method CHyQMOM, though the Monte Carlo moment errors dominate the QBMM ones and so CQMOM has the same results.
Indeed, the error converges at the expected rate.

\subsection{Bubble cavitation}

The dynamics of a cavitating gas bubble dispersion serves as a two-internal-coordinate nonlinear example problem.
The Rayleigh--Plesset equation models the bubble dynamics~\citep{brennen95}:
\begin{gather}
    R \Rddot + \frac{3}{2} \Rdot^2 + \frac{4}{\Rey} \frac{\Rdot}{R} = \frac{1}{R^3} - C_p,
    \label{e:rpe}
\end{gather}
where $R$ is the bubble radius, $\Rey$ is the Reynolds number (dimensionless ratio of inertial to viscous effects), and $C_p$ is the dimensionless pressure ratio between the suspending fluid and bubbles.
Thus, $R$ and $\Rdot$ are the two internal coordinates ($\bxi$).
For our purposes it suffices to ignore surface tension effects (following~\eqref{e:rpe}) and use $C_p = 1/0.3$ to represent a relatively large pressure ratio.
This formulation is non-dimensionalized by the (monodisperse) equilibrium bubble radius and suspending fluid density and pressure. 
The initial NDF is a log-normal distribution in the $R$-coordinate (shape parameters $\mu_R = 1, \sigma_R = 0.2$) and a normal distribution in the $\Rdot$ coordinate ($\mu_{\Rdot} = 0$, $\sigma_{\Rdot} = 0.1$).
The NDF is initially uncorrelated.

\begin{figure*}
    \centering
    \includegraphics{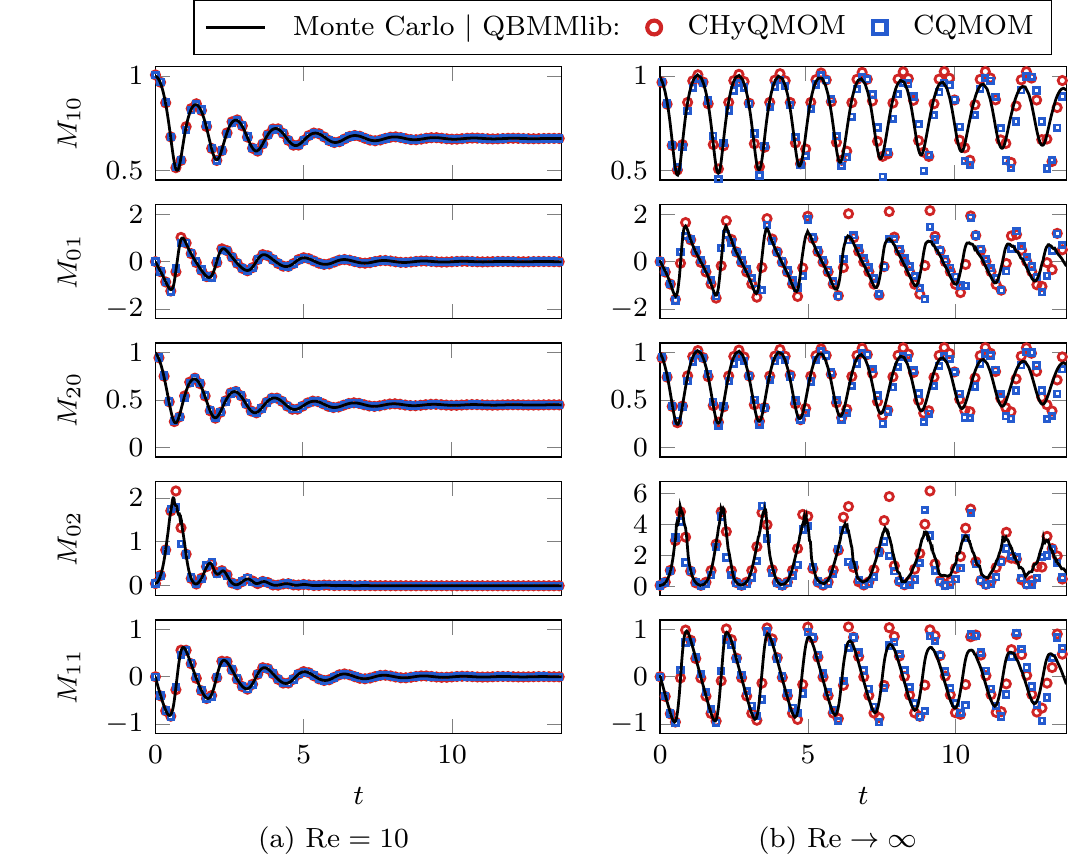}
    \caption{Evolution of the first- and second-order moments for (a) viscous and (b) inviscid bubble dynamics as labeled.
    Exact moments are approximated by a (sufficiently well converged) $N_\text{MC} = 5000$ Monte Carlo simulation.
    The symbols show QBMMlib solutions for $N_{\hxi_1} = N_{\hxi_2} = 2$ CHyQMOM and CQMOM, as labeled.
    }
    \label{f:bubmom}
\end{figure*}

Figure~\ref{f:bubmom} shows the moment dynamics for two bubble dispersions problems: (a) viscous  $\Rey = 10$ and (b) inviscid $\Rey \to \infty$.
Here, $\Rey = 10$ is the Reynolds number that corresponds to $\SI{1}{\micro\meter}$ bubbles in water and $\Rey \to \infty$ represents ignoring viscous effects.
Invoking $\Rey \to \infty$ is not appropriate for most cavitation problems of physical relevance, though it provides a useful reference.
In both cases the mean bubble radius $\mom_{10}$ oscillates and damps.
This damping is more significant in (a) than (b) due to viscous effects, as expected.
In the $\Rey = 10$ case, this is sufficient for the QBMMlib-predicted moments to match the Monte Carlo results.
However, for $\Rey \to \infty$, the evolving moment set is unable to faithfully represent the bubble oscillations, particularly at long times.
Indeed, a mismatch between the Monte Carlo and QBMMlib results is clear for $\mom_{02}$ and $\mom_{11}$.
These differences are qualitatively similar for both the CQMOM and CHyQMOM algorithms.
This is because closing the moment system requires extrapolating out of the represented moment space, which is of similar fidelity for both algorithms.

\section{Impact and conclusions}\label{s:impact}

This paper introduced QBMMlib, a library for solving PBEs using quadrature-based moment methods.
It is a Wolfram Language package, which is useful for automating the procedure of using QBMMs for simulating phenomena like bubble and particle dynamics.
This includes constructing a moment set for a given QBMM, determining the right-hand-side functions corresponding to a governing equation automatically, and inverting the moment set for quadrature points to close the system.
These routines leverage Mathematica's symbolic algebra features and include modern QMOM and conditional-QMOM methods.
Having these features available in a unified framework is helpful, particularly when it is unclear what QBMM will be appropriate (or stable) for the model dynamics.
Our searches suggest that QBMMlib is the only library, open source or otherwise, that provides such capabilities.
Given this, QBMMlib should help researchers prototyping QBMMs for their physical problems (or developing new QBMMs entirely).
Indeed, the authors used QBMMlib to guide the implementation of CHyQMOM for phase-averaged bubble cavitation into MFC, the first flow solver with this capability~\citep{zhang94,bryngelson19_CPC}.

\section*{Conflict of Interest}

We wish to confirm that there are no known conflicts of interest associated with this publication and there has been no significant financial support for this work that could have influenced its outcome.

\section*{Acknowledgements}

The authors appreciate the insights of Professor Alberto Passalacqua when developing this library.
The US Office of Naval Research supported this work under grant numbers N0014-17-1-2676 and N0014-18-1-2625.

\bibliographystyle{elsarticle-num-names} 
\bibliography{arxiv}

\end{document}